\documentclass[preprint,aps,superscriptaddress,showpacs]{revtex4}% Physical Review B

\usepackage{graphicx}% Include figure files
\usepackage{dcolumn}% Align table columns on decimal point
\usepackage{bm}% bold math
\usepackage{subfigure}
\usepackage{color}      % use if color is used in text
\usepackage{amsmath}      % use if color is used in text
\usepackage{hyperref}
\newcommand{ \ang}[0]{\AA$^3$}
\newcommand{ \mub}[0]{$\mu_{\mathrm{B}}$}

\newcommand{\totfig}[1]{(Color online) Total energy (a) and spin magnetic moments (b) for #1 as obtained with the VASP code for the SCAN (red), PBE (blue) and LSDA (black) functionals. Vertical and horisontal dashed lines correspond to experimental values given in Tab.\ \ref{tab:summary_vasp}.}

\newcommand{\totfighcp}[1]{(Color online) Total energy (a) and spin magnetic moments (b) for #1 as obtained with the VASP code for the SCAN (red), PBE (blue) and LSDA (black) functionals. Vertical dashed lines correspond to the experimental volume given in Tab.\ \ref{tab:summary_vasp}.}

\newcommand{\dosfig}[1]{(Color online) Spin-resolved DOS for #1 calculated at the experimental volume with the VASP code for the SCAN (blue), PBE (red), and LSDA (black) functionals, where $E_F$ is the Fermi energy.}

\newcommand{\ifm}[0]{Department of Physics, Chemistry and Biology (IFM)
Link\"oping University, SE-58183 Link\"oping, Sweden}

\newcommand{\serc}[0]{Swedish e-Science Research Centre (SeRC), Link\"oping University, SE-58183 Link\"oping, Sweden
}

\newcommand{\misis}[0]{Materials Modeling and Development Laboratory, National University of Science and Technology ''MISIS'', 119049 Moscow, Russia
}

\newcommand{\mpie}[0]{Max-Planck-Institut f\"ur Eisenforschung GmbH, Max-Planck Strasse 1, 40237 D\"usseldorf, Germany
}

\date{\today}

\begin{document}
\title{Assessing the SCAN functional for itinerant electron ferromagnets}

\author{M.~Ekholm}
\email{marekh@ifm.liu.se}
\affiliation{\ifm}
\affiliation{\serc}

\author{D.~Gambino}
\affiliation{\ifm}

\author{H.~J.~M.~J\"onsson}
\affiliation{\ifm}

\author{F.~Tasn\'adi}
\affiliation{\ifm}

\author{B.~Alling}
\affiliation{\ifm}
\affiliation{\mpie}

\author{I.~A.~Abrikosov}
\affiliation{\ifm}
\affiliation{\misis}


\begin{abstract}
Density functional theory is a standard model for condensed matter theory and computational material science.
The accuracy of density functional theory is limited by accuracy of the employed approximation to the exchange-correlation functional.
Recently, the so-called strongly constrained approprietly normed (SCAN) \cite{sun15} functional has received a lot of attention due to promising results for covalent, metallic, ionic, as well as hydrogen- and van der Waals-bonded systems alike.
In this work we focus on assessing the performance of the SCAN functional for itinerant magnets by calculating basic structural and magnetic properties of the transition metals Fe, Co and Ni.
We find that although structural properties of bcc-Fe seem to be in good agreement with experiment, SCAN performs worse than standard local and semilocal functionals for fcc-Ni and hcp-Co.
In all three cases, the magnetic moment is significantly overestimated by SCAN, and the $3d$ states are shifted to lower energies, as compared to experiments.
\end{abstract}

\def \mysize{0.5}

\maketitle

\section{Introduction}
\label{sec:intro}
\noindent
Transition metals, and their alloys, make up the backbone of materials that are associated with the industrial age.
Among the transition metals, only Fe, Co and Ni display ferromagnetism at ambient conditions.
In these metals, $3d$ electrons collectively form localised magnetic moments due to Pauli exchange and Coulomb correlation.
However, the same $3d$ electrons also take part in the chemical bonding, and may therefore be classified as itinerant. \cite{mott64}
\textit{Ab-initio} theory must correctly account for the dual nature of these electrons, itinerant and localised, which will affect not only the magnetic moment but also the bond lengths.

Density-functional theory\cite{hohenberg64,kohn65} (DFT) has provided the basis for the theory of itinerant electron magnetism.\cite{kublerbook}
A limiting factor to the accuracy of DFT calculations is the employed approximation of exchange- and correlation-effects.
The simplest of its kind is the local spin density approximation\cite{kohn65,perdew92} (LSDA), which only takes the local spin polarised density as input.
However, in $3d$-systems, a major drawback of the LSDA is the inherent overbinding, which may lead to wrong conclusions as regards magnetism.
For instance, LSDA predicts non-magnetic fcc-Fe to be lower in energy than the correct ferromagnetic bcc phase at their respective equilibrium volumes.\cite{kublerbook}
This failure may be ascribed to overbinding, since at the larger, experimental, volume LSDA correctly favours the ferromagnetic bcc phase.\cite{suppmat}

The generalised gradient-approximation (GGA) goes beyond LSDA by including also the local gradient of the density, making them semilocal.
In bcc-Fe, GGA parametrisations such as PW91\cite{perdew92} or PBE\cite{perdew96} improves the equilibrium volume, and thereby reproduce the ferromagnetic bcc phase as the ground state self-consistently.
Yet, there are examples where the above standard GGAs still fail to correct the magnetic picture, such as the Mn-rich side of FeMn phase diagram.\cite{ekholm11} 

More recent functionals on GGA-form, such as PBEsol\cite{perdew08} and AM05\cite{armiento05}, have been shown to improve bonding properties as compared to PBE in many cases.\cite{ropo08,haas09}
However, these functionals do not improve the equilibrium volume of magnetic transition metals, yielding values in between those of LSDA and PBE, which typically is an underestimation compared to experiment.
These elemental systems thus remain challenging for \textit{ab-initio} theory, and the description of correlations in their electronic structure at ambient conditions is an area of active research.\cite{irkhin93,schafer05,katanin10,sanchez12,sponza17}

So-called meta-GGAs also take into account the kinetic energy density of the non-interacting electrons.
The recently proposed non-empirical strongly constrained approprietly normed (SCAN) \cite{sun15} meta-GGA functional has been demonstrated to improve on the standard GGAs in several systems with various types of bonding.\cite{kitchaev16,patra17,shahi18,furness18,sun16,hinuma17,zhang17,zhang18}
By virtue of being effectively semilocal, the computational cost of SCAN is comparable to regular GGAs, which makes it especially attractive for the high-throughput approach to computational materials science.

However, there are few studies focusing on the performance of SCAN for magnetic systems.
For the magnetic insulator BiFeO$_3$, where Sun et al.\ found the local Fe magnetic moment closer to the experimental value with SCAN as compared to PBE.\cite{sun16}
Very recently, Isaacs and Wolverton \cite{isaacs18} tested SCAN for a large set of systems, which also included spin magnetic moments for Fe, Co and Ni.
SCAN was deemed an overall improvement over PBE for magnetic systems, although the magnetic moments were larger than the experimental atomic magnetic moments in the ferromagnetic transition metals.
Similar results for bcc-Fe and fcc-Ni have also appeared in a manuscript by Jana et al.\ \cite{jana18}.
In this work, we assess the SCAN functional for the ferromagnetic metals bcc-Fe, hcp-Co and fcc-Ni, with an emphasis on structural parameters, electronic structure, and the role of spin-orbit coupling.
In Sec.\ \ref{sec:methods} we give details of the computational method.
We then present results for the three systems separately in Sec.\ \ref{sec:results}, before discussing general conclusions in Sec.\ \ref{sec:summary}.

\section{Methods}
\label{sec:methods}
We have performed scalar-relativistic calculations of total energy, density of states (DOS) as spin-orbit coupling is not expected to significantly affect structural properties in these systems.\cite{lejaeghere14}
The magnetic moment has been evaluated both with and without spin-orbit coupling.
In all cases, the reported spin moments correspond to the total spin magnetization of the unit cell.
Our main computational method is the projected augmented waves method (PAW)\cite{blochl94} as implemented in the Vienna ab-initio simulation package (VASP)\cite{vasp2,vasp3}, version 5.4.4.
We used the PAW Fe\_sv, Ni\_pv, and Co\_sv potentials with the plane wave cut-off energy to 600 eV. 
A $31\times 31 \times 31$ k-point grid was used from which special k-points were chosen with the Monkhorst-Pack\cite{monkhorst76} scheme.
Total energy and DOS was evaluated using the Bl\"ochl tetrahedron method.\cite{tetrahedron}

Auxiliary calculations were also made using the all-electron full-potential (linearised) augmented plane waves plus local orbitals method of the Wien2k code,\cite{wien2k} version 17.1.
The muffin-tin radii were assigned the constant values $R_{\textrm{MT}}=1.60$ a.u.\ and 1.85 a.u.\ for bcc-Fe and fcc-Ni, respectively.
We used a $35\times 35 \times 35$ k-point grid in the full Brillouin zone for bcc-Fe and $36\times 36 \times 36$ for fcc-Ni.
In all calculations we set $R_{\textrm{MT}}K_{\textrm{max}}=10$.

We have also used the Quantum Espresso code\cite{qe1,qe2} for bcc-Fe, with the 
norm-conserving pseudopotential pbe-sp-mt\_gipaw.UPF, including semicore states in the valence, local part $d$, including gipaw reconstruction and two gipaw projectors per $p$ and $d$ channels, and only one projector on the $s$ channel.
The plane wave energy cut-off was 200 Ry and the k-mesh $22\times 22 \times 22$.

Total energy, $E$, was fitted to the third order Birch-Murnaghan equation of state: \cite{birch47}
\begin{widetext}
\begin{equation}
E(V) =  E_0 + \dfrac{9}{16}B_0 V_0  \left( 
 \left[ \left( \dfrac{V_0}{V} \right)^{2/3} - 1 \right]^3 B_0 '   
  + \left[ \left( \dfrac{V_0}{V} \right)^{2/3} - 1 \right]^2
\left[ 4\left( \dfrac{V_0}{V} \right)^{2/3} - 6 \right]
\right)
\end{equation}
\end{widetext}
where $E_0$ is the minimum energy, which is assumed at the equilibrium volume, $V_0$, and the bulk modulus, $B_0$, is defined as
\begin{equation}
B_0 = V \dfrac{d ^2 }{d V^2} E(V) \,  \,\Big\rvert _{V = V_0}
\end{equation}
and finally:
\begin{equation}
B_0 '  = - \dfrac{V}{B}  \dfrac{d  }{d V} B(V) \,\Big\rvert _{V = V_0}
\end{equation}

\section{Results}
\label{sec:results}
Tab.\ \ref{tab:summary_vasp} summarises our results based on VASP calculations in comparison with experimental values.
Our calculations have been performed on static lattices, without zero-point motion.
The experimental values for $V_0$, $B_0$ and $B_0'$ have therefore been collected from Ref.\ \cite{lejaeghere14}, which are adjusted for zero-point motion within the Debye model.
Experimental values for the spin magnetic moment, $m_s$ have been taken from Refs.\ \cite{wijn97,bonnenberg}
Below we discuss our results in detail for each element separately.
\begin{table}
\caption{
Ground state parameters calculated with the VASP code, and experimental values.
The experimental values of $V_0$, $B_0$, $B_0 '$ are taken from Ref.\ \cite{lejaeghere14},  where $V_0$ and $B_0$ were adjusted for zero-point motion.
Experimental values for the spin magnetic moment, $m_s$ have been collected from Refs.\ \cite{wijn97,bonnenberg}.
\label{tab:summary_vasp}
}
\begin{ruledtabular}
\begin{tabular}{l l  l l l l }
  		&  & 			$V_0$ 	& $B_0$ 	& $B_0'$  & $m_s(V_0)$ \\
  		 & &     [\AA$^3$ / atom] & [GPa] & [1] & [$\mu_{\mathrm{B}} $] \\
\hline
   & 	Exp.  & 		11.64 	    &	175.1       & 4.6  &  2.13  \\

& SCAN & 11.58 & 157.5 & 5.05 & 2.66 \\

bcc-Fe & PBE & 11.35 & 197.7 & 4.45 & 2.20 \\

& LSDA & 10.36 & 253.3 & 4.39 & 1.95 \\

\hline
 & Exp. &		10.81	    & 	  192.5         & 4 &  0.57 \\

& SCAN & 10.38 & 230.5 & 4.79 & 0.73 \\

fcc-Ni & PBE & 10.90 & 199.8 & 4.76 & 0.63 \\

& LSDA & 10.06 & 253.6 & 4.77 & 0.58 \\

\hline
  & Exp.	&		10.96	&  198.4      & 4.26  & -\\
  
& SCAN & 10.45 & 262.5 & 4.15 & 1.73 \\

hcp-Co & PBE & 10.91 & 196.9 & 4.61 & 1.61 \\

& LSDA & 9.99 & 237.6 & 4.95 & 1.49 \\

\end{tabular}
\end{ruledtabular}
\end{table}

\begin{table}
\caption{
Structural equilibrium properties calculated with the Wien2k code.
\label{tab:summary_wien2k}
}
\begin{ruledtabular}
\begin{tabular}{l l  l l l  }
  % & \multicolumn{4}{c}{$V_0$} \\
  		&  & 			$V_0$ 	& $B_0$ 	& $B_0'$  \\
  		 & &     [\AA$^3$ / atom] &  [GPa] & [1]  \\
\hline
 bcc-Fe              & SCAN &       11.13   & 204.1 & 5.01 \\
& PBE & 11.38 & 202.5 & 4.58    \\
\hline
 fcc-Ni          & SCAN   &    10.34  & 230.0 & 4.87  \\     
 & PBE     &      10.89    &  200.9    & 4.76    \\
\hline
\end{tabular}
\end{ruledtabular}
\end{table}

\subsection{bcc-Fe}
\begin{figure}
\begin{center}
\subfigure[]{
\includegraphics[scale=\mysize]{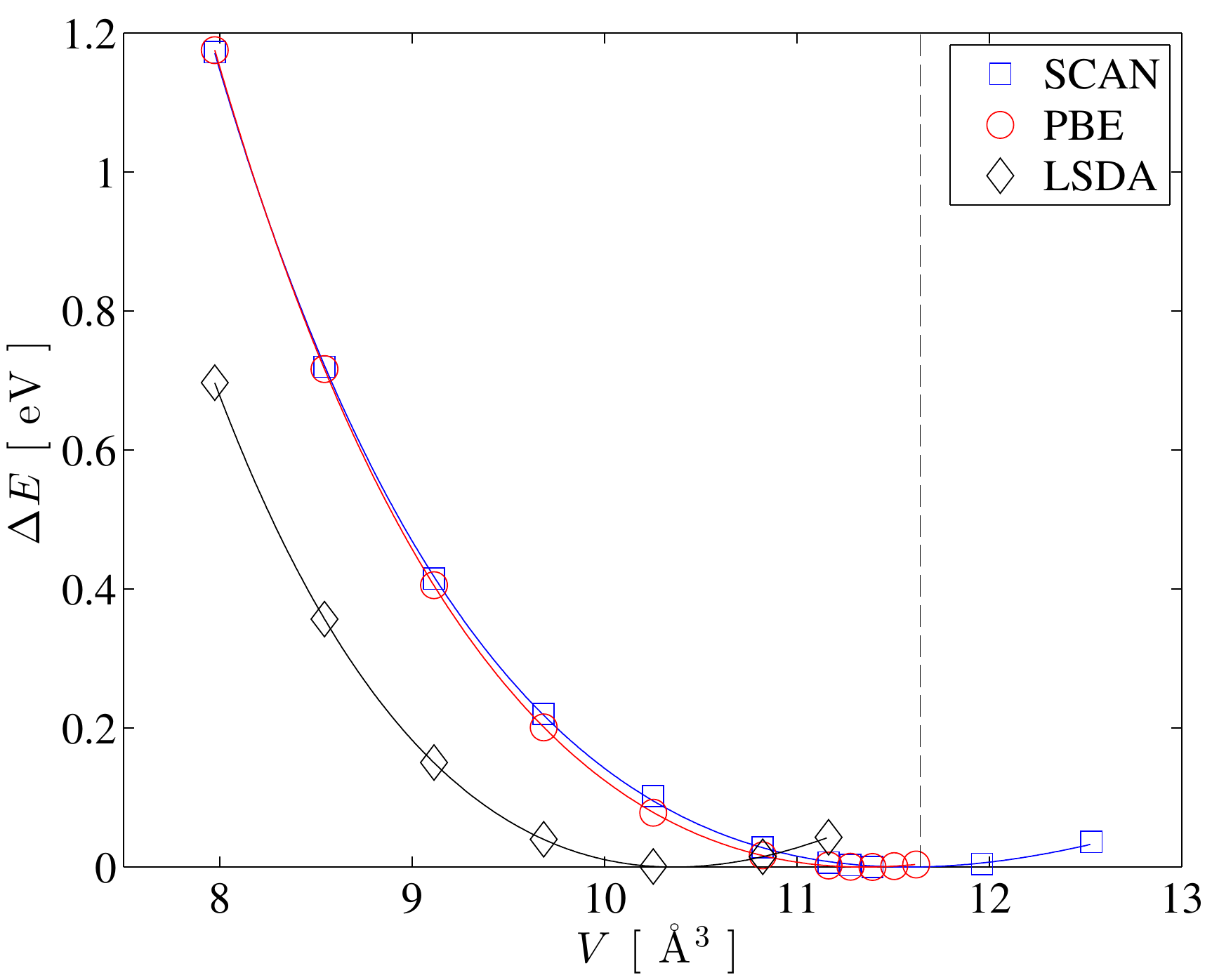}
}
\subfigure[]{
\includegraphics[scale=\mysize]{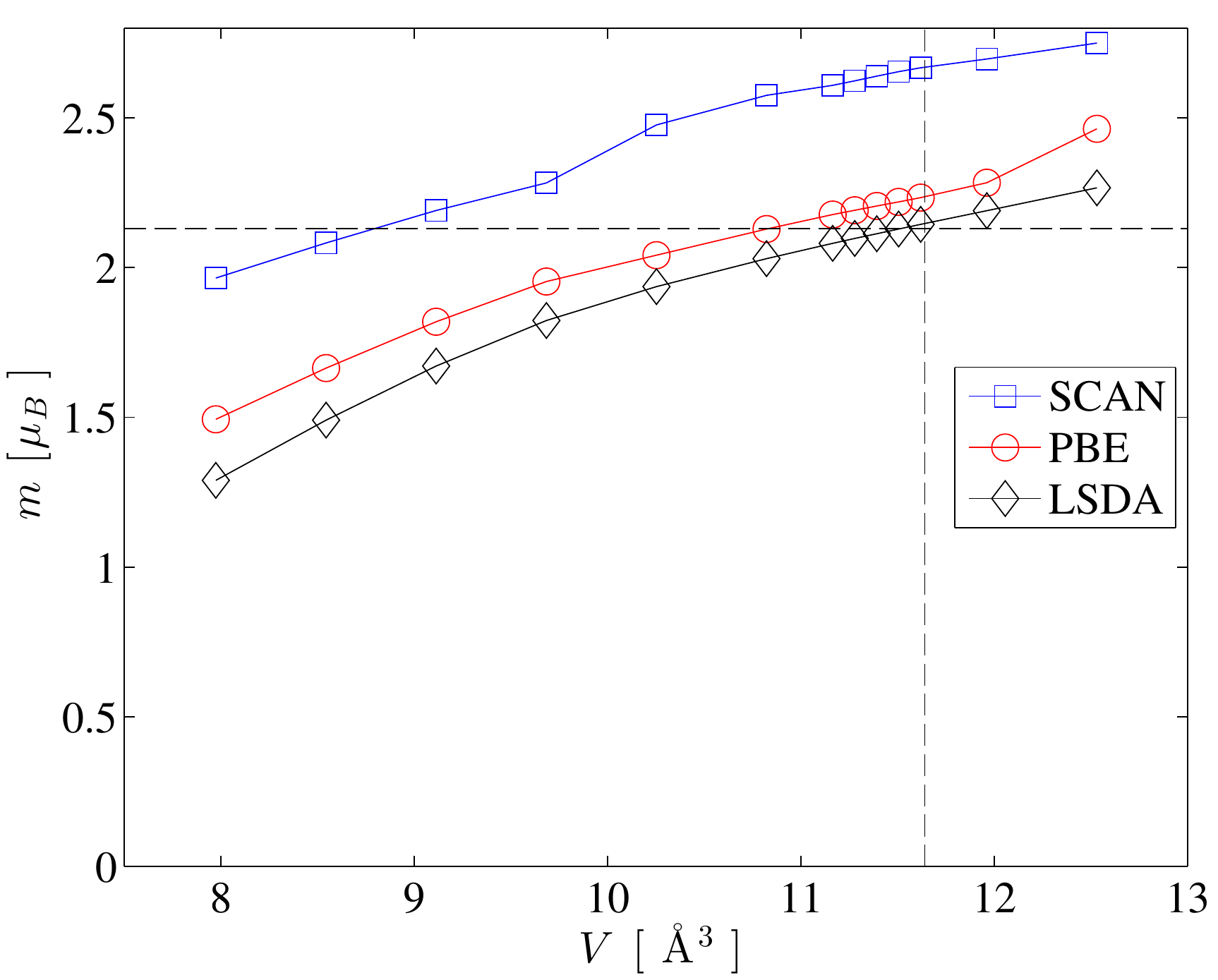}
}
\end{center}
\caption{\totfig{bcc-Fe}
\label{fig:bccFeVASP}
}
\end{figure}

In Fig.\ \ref{fig:bccFeVASP} we show total energy and spin magnetic moments as a function of volume for bcc-Fe.
As discussed in Sec.\ \ref{sec:intro}, LSDA overbinds, which results in an underestimated equilibrium volume, 10.36 \ang\ compared to the experimental value of 11.68 \ang.
The corresponding bulk modulus is overestimated at 253.3 GPa compared to experiment (175.1 GPa).
This is partially corrected by PBE, which yields the equilibrium volume 11.38 \ang\ with a reduced bulk modulus of 197.7 GPa.
With the SCAN functional we obtain a yet larger volume than with PBE (11.58 \ang), which is closer to the experimental value and a lower bulk modulus (157.5 GPa).

\begin{figure}
\includegraphics[width=9.5cm]{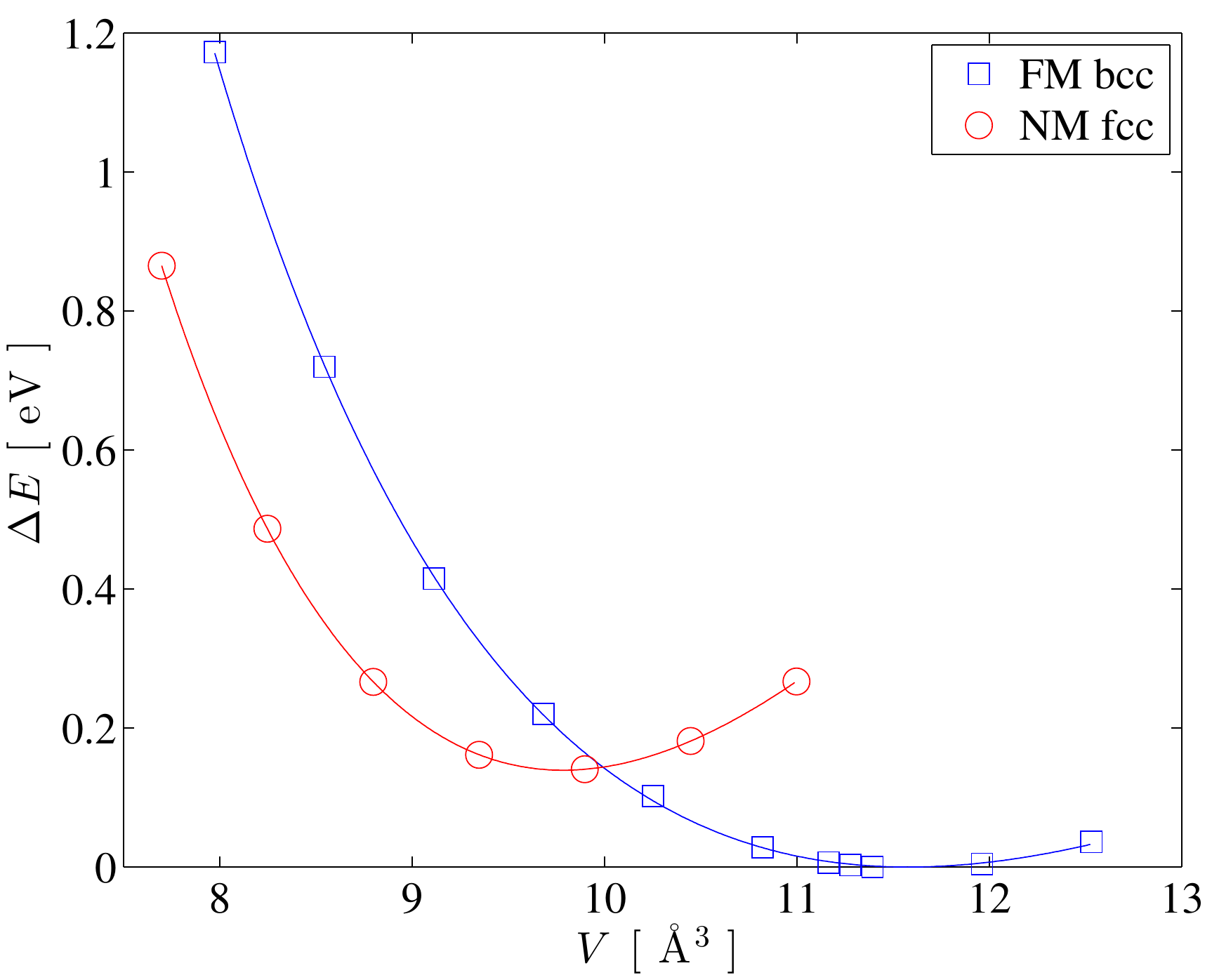}
\caption{Total energy for non-magnetic (NM) fcc-Fe and ferromagnetic (FM) bcc-Fe  calculated with the VASP code using the SCAN functional.
\label{fig:fccFe_SCAN}}
\end{figure}

However, SCAN calculations with the all-electron Wien2k code seem to yield conflicting results.
Instead of an increased equilibrium volume as compared to PBE, we obtain a smaller volume (11.13 \ang) with a similar bulk modulus of 204.1 GPa (Tab.\ \ref{tab:summary_wien2k}).
This discrepancy may be due to the non self-consistent implementation\cite{tran16} of the SCAN functional in Wien2k.
Calculations with the norm-conserving potentials of the Quantum Espresso code (Tab.\ \ref{tab:summary_qe}) support the notion of a larger volume with SCAN (11.55 \ang) than with PBE (11.24 \ang).
We also note that SCAN self-consistently recovers bcc-Fe as more stable than the nonmagnetic fcc-solution, as shown in Fig.~\ref{fig:fccFe_SCAN}.

At any given volume, we find the spin magnetic moment produced by SCAN larger than that of PBE, which in turn is always larger than that of LSDA (see Fig.\ \ref{fig:bccFeVASP} b).
At the respective theoretical equilibrium volumes, the SCAN spin moment is severely overestimated, 2.66 \mub, compared to the experimental value of 2.13 \mub.
This effect is reminiscent of results for BiFeO$_3$, where the local Fe magnetic moment was also seen to be overestimated.\cite{sun16}
Nevertheless, in that system this was still considered an improvement over PBE.
The PBE spin moment is closest to the experimental value when evaluated at the theoretical equilibrium volume.
However, it should be noted that at the experimental volume, the LSDA spin moment in fact gives the best agreement with experiment of the considered functionals.
Tab.\ \ref{tab:summary_so} shows also orbital magnetic moments calculated at the experimental volume.
Adding the spin and orbital moments allows the total moment to be compared to the experimental value for the saturation magnetisation of 2.21 \mub\.
Although both PBE and SCAN underestimate the orbital moment, the atomic magnetic moment is still significantly overestimated compared to the experimental value.

\begin{figure}
\begin{center}
\includegraphics[scale=\mysize]{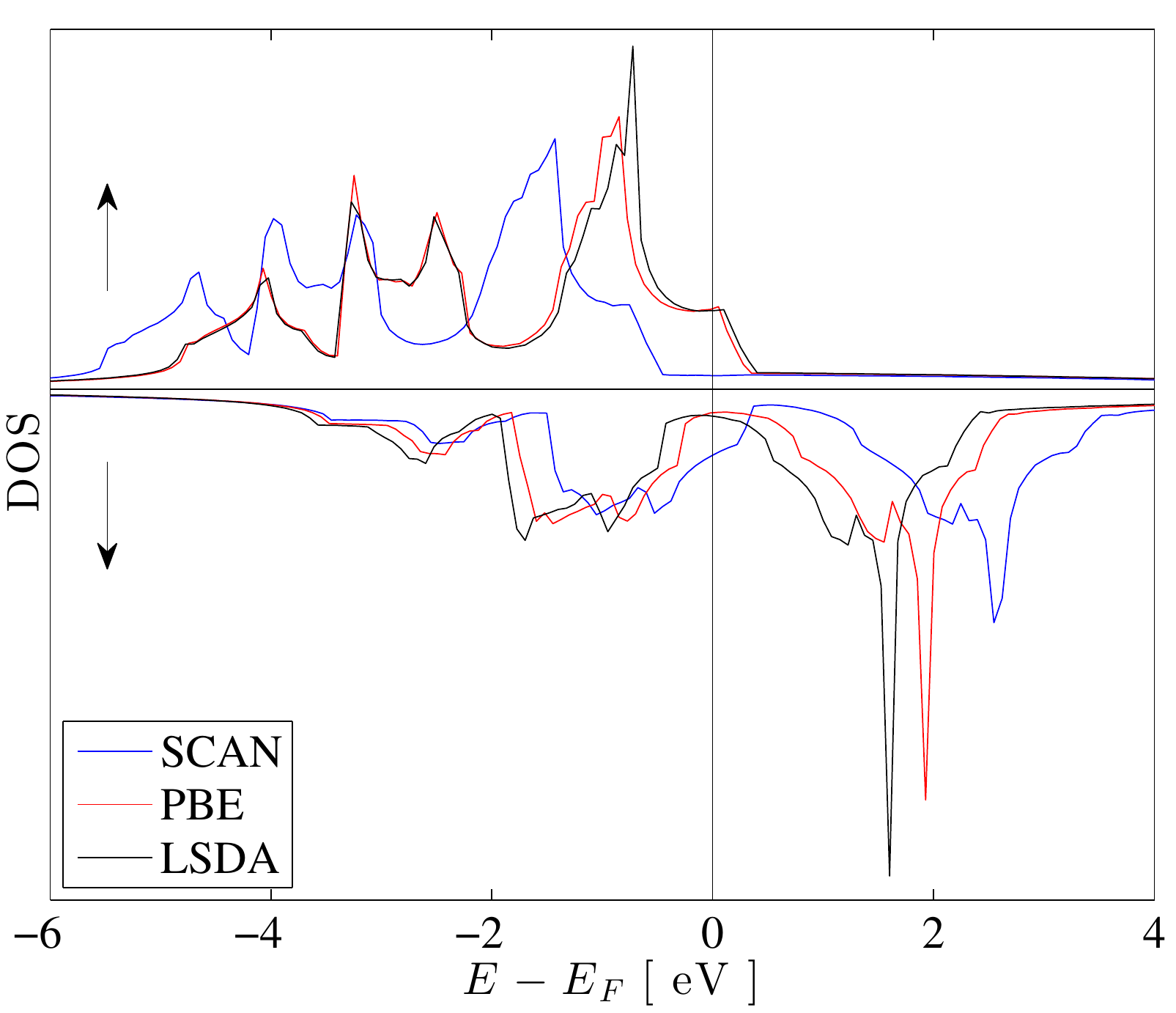}
\end{center}
\caption{\dosfig{bcc-Fe}
\label{fig:bccFeexpdos}
}
\end{figure}

In order to evaluate the impact of the SCAN functional on the electronic structure, we compare all three functionals at the experimental volume, which is shown in Fig.~\ref{fig:bccFeexpdos}. 
A similar comparison at the respective equilibrium volume calculated with each functional is shown in the Supplementary Material \cite{suppmat}.
Spin-up bands are here identified with the majority spin carriers.
The large magnetic moment obtained with SCAN is reflected in a large exchange-splitting of the $d$-bands.
bcc-Fe is usually considered a weak ferromagnet, in the sense that the spin-up band is not completely filled.
This is indeed in line with LSDA and PBE-calculations, which position the Fermi level in a minimum in the spin-down DOS, which in turn determines the spin splitting.
However, in the SCAN picture it seems more favourable to fill the spin-up bands, making bcc-Fe a strong ferromagnet with a very large magnetic moment.

\begin{table}
\caption{
Equilibrium properties calculated with the Quantum Espresso code.
\label{tab:summary_qe}
}
\begin{ruledtabular}
\begin{tabular}{l l  l l l l }
  		&  & 			$V_0$ 	& $B_0$ 	& $B_0'$ &  $m_s$ \\
  		 & &    [ \AA$^3$ / atom ]& [GPa ]& [1]  &[ $\mu_{\mathrm{B}} $] \\
\hline
bcc-Fe & SCAN & 11.55& 181.5 & 6.4 & 2.74  \\
        & PBE &       11.24   & 190.8 & 4.99 & 1.82\\
\end{tabular}
\end{ruledtabular}
\end{table}

\begin{table}
\caption{
Spin and orbital magnetic moments from calculations including spin-orbit coupling with the  VASP code. For each element, the volume was set to the respective experimental value, as stated in Tab.~\ref{tab:summary_vasp}.
The quantisation axis were taken as [100] for bcc-Fe, [111] for fcc-Ni and [001] for hcp-Co.
Values in parenthesis correspond to the spin moments obtained without spin-orbit coupling.
Experimental values for Fe and Ni are taken from Ref.\ \cite{wijn97}, and \cite{stearns} for hcp-Co.
\label{tab:summary_so}
}
\begin{ruledtabular}
\begin{tabular}{l l  l l l l }
  		&  & 			$m_s$ 	& $m_o$ 	& $m_{\mathrm{tot}}$  \\
  		 & &   [ \mub\ ] & [ \mub\ ]& [ \mub\ ]  \\
\hline
bcc-Fe & SCAN         & 2.65 (2.66)  & 0.034     &  2.68   \\
              & PBE             & 2.22 (2.22) &  0.042    &  2.26        \\
              & Exp              & 2.13 &  0.080    &    2.21   \\     
 fcc-Ni &     SCAN      &  0.745 (0.745) &   0.059  &  0.804 \\
              &   PBE          &   0.629 (0.631)           & 0.048 & 0.677   \\
        & Exp                  &     0.57                &   0.050    &  0.62      \\     
hcp-Co & SCAN         & 1.76 (1.76)                  & 0.084     &  1.84     \\
              & PBE             &  1.61  (1.61)                 &  0.075    &  1.69         \\
              & Exp              &            -     & -               &    1.715--1.728         \\   
\end{tabular}
\end{ruledtabular}
\end{table}

\subsection{fcc-Ni}

\begin{figure}
\begin{center}
\subfigure[]{
\includegraphics[scale=\mysize]{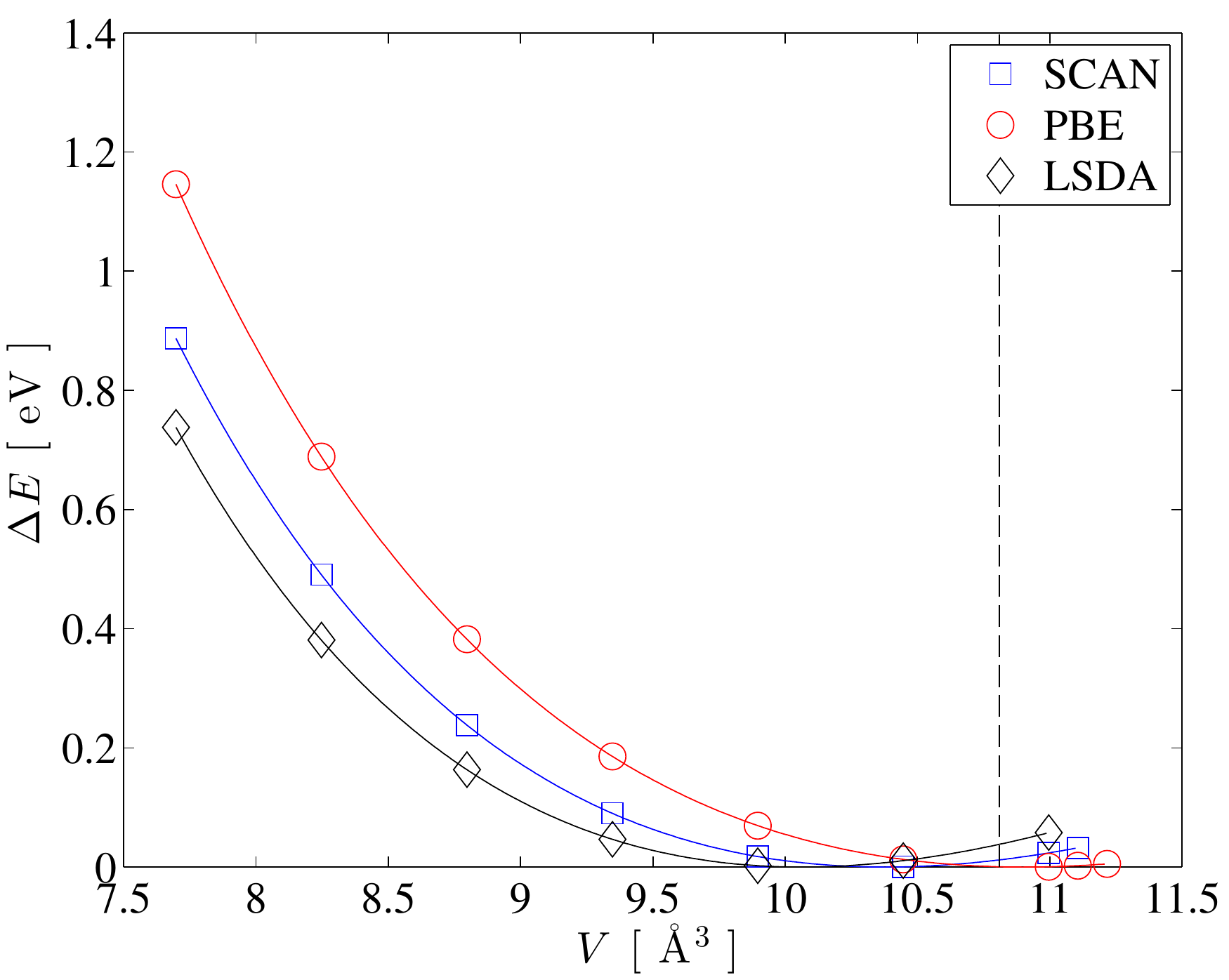}
}
\subfigure[]{
\includegraphics[scale=\mysize]{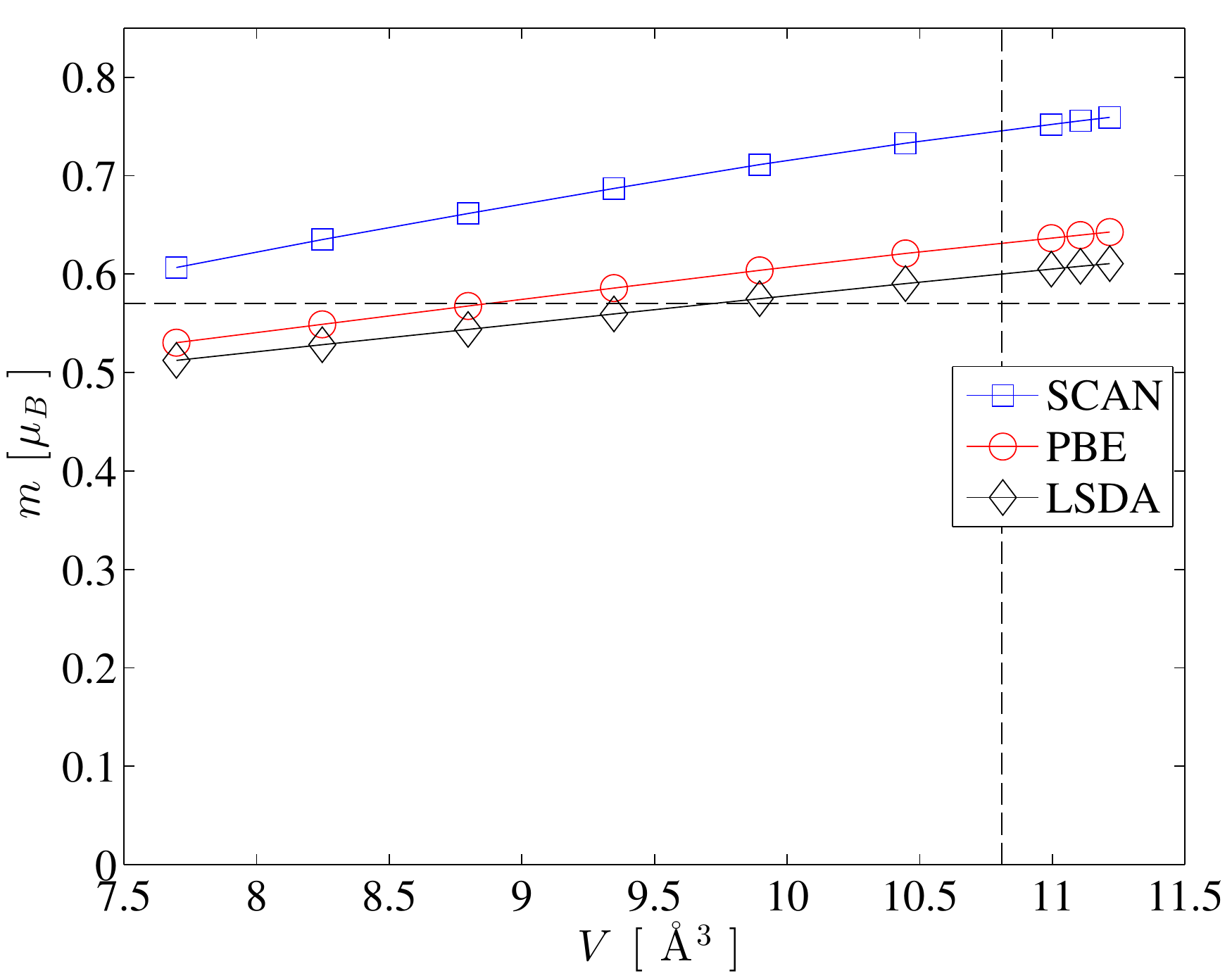}
}
\caption{\totfig{fcc-Ni}
\label{fig:fccNiVASP}
}
\end{center}
\end{figure}

The total energy curves obtained with VASP for fcc-Ni are shown in Fig.\ \ref{fig:fccNiVASP}, and equilibrium properties are listed in Tab.\ \ref{tab:summary_vasp}.
The results are in very good agreement with our Wien2k calculations (see Tab.\ \ref{tab:summary_wien2k}) and values published in Ref.\ \cite{tran16}.
In contrast to bcc-Fe, SCAN seems to strengthen the bonds as compared to PBE, with an equilibrium volume of 10.38 \ang, as compared to 10.90 \ang.
The bulk modulus is also increased with SCAN (230.5 GPa) as compared with PBE (199.8 GPa).
This correction by SCAN is in the right direction, since the PBE seems to produce too soft bonds in fcc-Ni.
However, the effect is exaggerated, making the agreement with experiment clearly worse than for PBE.

The spin magnetic moment shown in Fig.\ \ref{fig:fccNiVASP} is seen to be consistently larger with SCAN than PBE.
At the equilibrium volume, the SCAN spin moment (0.73 \mub) is still much larger than the experimental value of 0.57 \mub.
All three functionals show a tendency to overestimate the magnetic moment.
LSDA therefore comes closest to the experimental spin moment at both the theoretical and experimental volumes.
Adding the orbital moment (Tab.\ \ref{tab:summary_so}) and comparing the total moment to the saturation magnetisation does not change the picture, as both PBE and SCAN produce orbital moments in good agreement with experiment.

\begin{figure}
\begin{center}
\includegraphics[scale=\mysize]{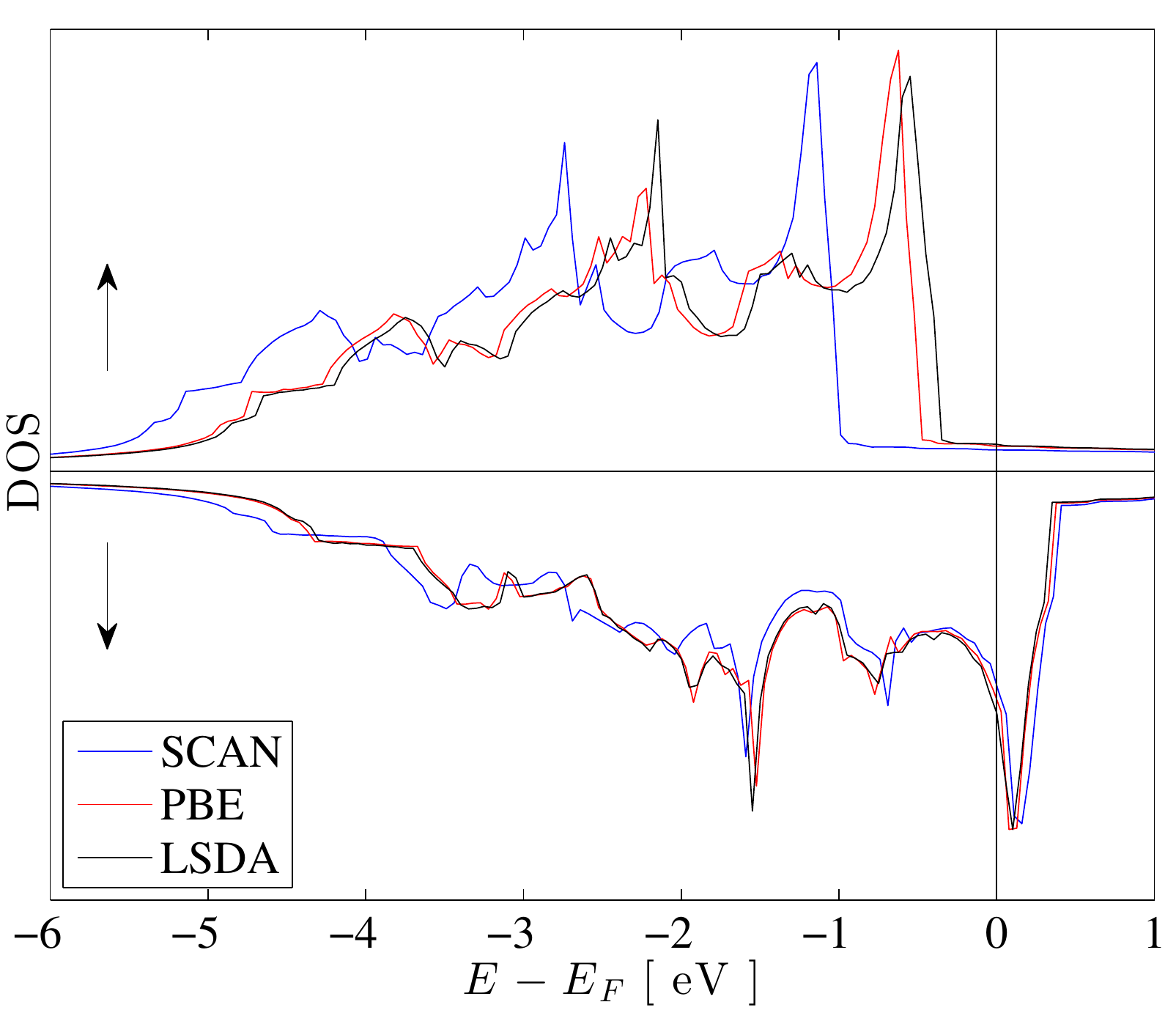}
\end{center}
\caption{\dosfig{fcc-Ni}
\label{fig:fccNiDOS}
}
\end{figure}

The DOS, evaluated at the experimental volume, is shown in Fig.\ \ref{fig:fccNiDOS}.
In the LSDA/PBE-picture the spin-up $d$-bands are virtually filled, and SCAN redistributes approximately $0.1$ $d$-electrons from the spin-down to the spin-up bands.
The resulting change in the spin magnetic moment is just below 0.2 \mub, which however is a change of roughly 30\%, due to the small spin moment of Ni.
Due to the more close packed crystal structure of fcc as compared to bcc, the canonical band structure shows less pronounced minima and maxima, which means that the modifications to the spin-down band appear more evenly distributed, than in bcc-Fe.

\subsection{hcp-Co}

\begin{figure}
\begin{center}
\subfigure[\label{fig:hcpCoVASPtote}]{
\includegraphics[scale=\mysize]{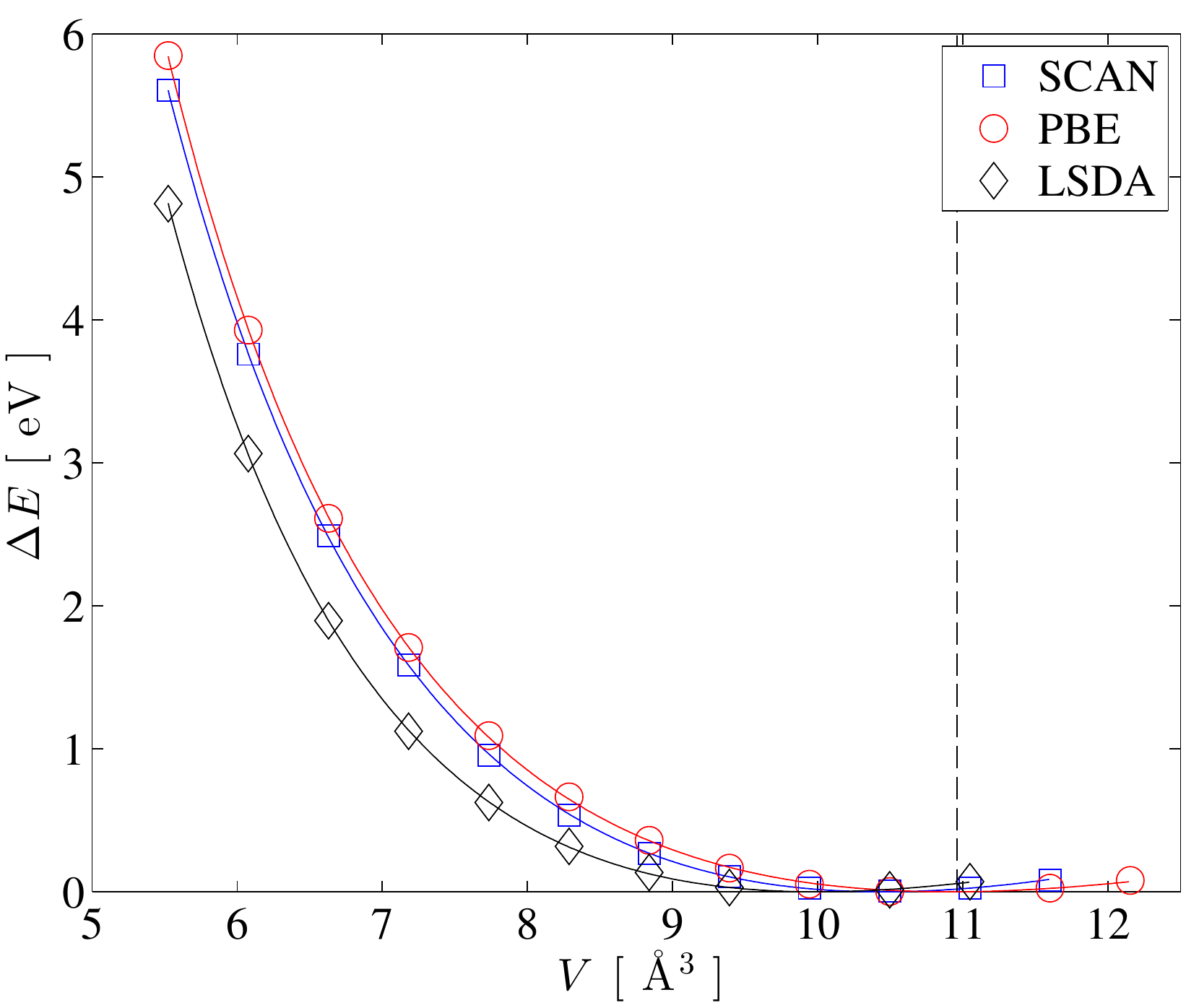}
}
\subfigure[\label{fig:hcpCoVASPmag}]{
\includegraphics[scale=0.42]{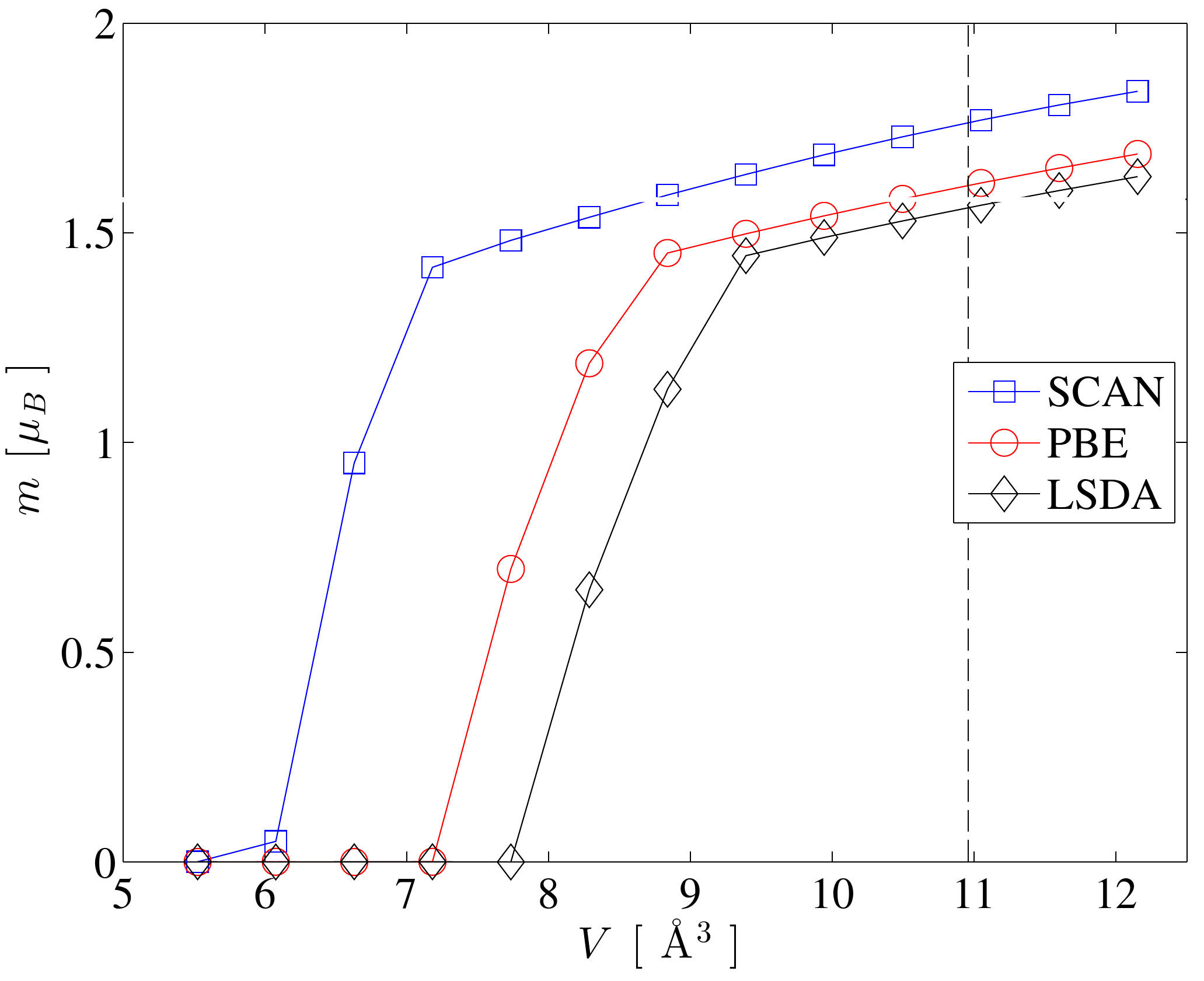}
}
\end{center}
\caption{{\footnotesize \totfighcp{hcp-Co} In (b), the $c/a$ ratio was set to 1.62. }
\label{fig:hcpCoVASP}}
\end{figure}

\begin{figure}
\begin{center}
\includegraphics[scale=\mysize]{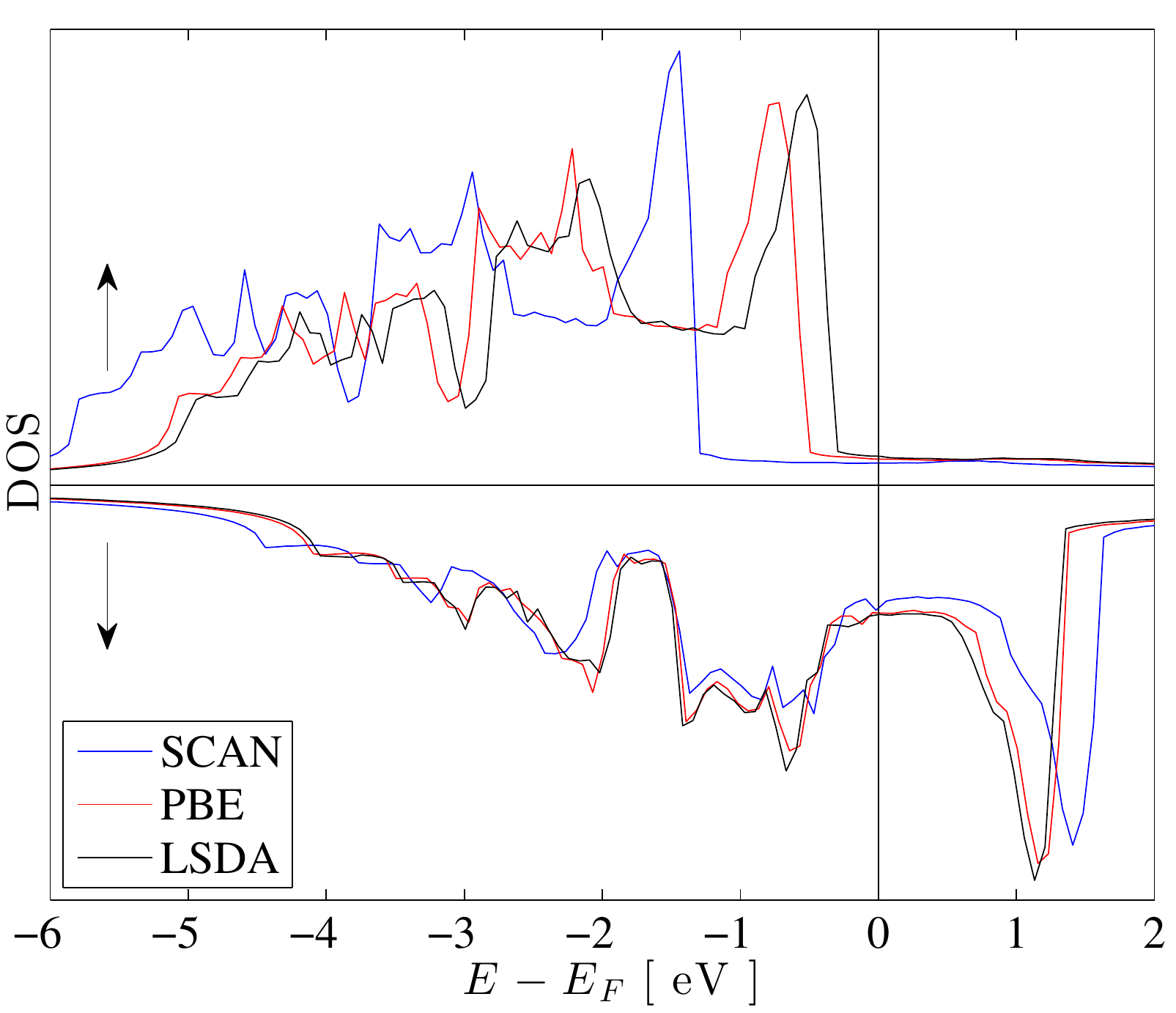}
\end{center}
\caption{\dosfig{hcp-Co}
\label{fig:hcpCoDOS}
}
\end{figure}

Fig.\ \ref{fig:hcpCoVASPtote} shows total energy as a function of volume for hcp-Co, as obtained with VASP.
At each volume, we optimised the $c/a$-ratio, which was found to be between 1.619 and 1.628 for all functionals.
SCAN produces a smaller equilibrium volume than PBE, 10.45 \ang, as compared with 10.91 \ang (see Tab.\ \ref{tab:summary_vasp}).
These values are both smaller than experiment, and the SCAN bulk modulus is also seen to be severely overestimated (262.5 GPa). 
In fact, this value of the bulk modulus is larger than that of LSDA (237.6 GPa), although the LSDA equilibrium volume is smaller (9.99 \ang).

The spin magnetic moment produced by LSDA at the theoretical equilibrium volume is 1.49 \mub, while 
PBE and SCAN yield 1.61 \mub\ and 1.73 \mub, respectively.
We are not aware of any available experimental data for low temperature spin- and orbital magnetic moments in hcp-Co, which makes the performance for the spin moment difficult to evaluate.
The total moment of the hcp-phase has been reported as 1.715--1.728 \mub \cite{stearns}, which is slightly smaller than the magnetic moment of the fcc-phase of 1.75 \mub \cite{crangle55}.
In our calculations, we obtain an orbital contribution to the magnetic moment of 0.08 \mub (Tab.~\ref{tab:summary_so}).
This means that SCAN again significantly overestimates the atomic magnetic moment.

Fig.\ \ref{fig:hcpCoDOS} shows the DOS calculated at the experimental volume with $c/a=1.62$.
LSDA and PBE results in filled spin-up bands.
Comparing to PBE, SCAN redistributes 0.05 electrons from the spin-down to the spin-up bands, which increases the spin magnetic moment by 0.1 \mub.

\section{Summary and conclusions}
\label{sec:summary}
We have calculated equilibrium properties of Fe, Co and Ni with the SCAN exchange-correlation functional.
The performance of SCAN, in comparison with LSDA and PBE, seems to follow similar trends in fcc-Ni and hcp-Co. 
In these systems, SCAN overbinds, resulting in a reduced equilibrium volume and overestimated bulk modulus.
This in contrast to the case of bcc-Fe, where SCAN results in an only slightly overestimated equilibrium volume and lower bulk modulus, improving on PBE results.

Nevertheless, the magnetic moment is severely overestimated by SCAN in all three systems.
At the experimental volume, the most accurate spin magnetic moments are still obtained with LSDA.
The large spin moments in SCAN is seen to arise from an increased exchange-splitting compared to LSDA and PBE.
In bcc-Fe, the spin-up bands even become filled, so that it goes from a weak to a strong ferromagnet.
This also means that the band width is larger with SCAN compared to PBE.
It is well known that the energy states produced by PBE and LSDA are too far from the Fermi energy compared to the experimental electron spectra.\cite{schafer05,sanchez09,sanchez12} 
There is no formal justification to identify Kohn-Sham eigenvalues with excitation energies.\cite{almbladh85} 
Nevertheless, it should be noted that the increased exchange splitting of SCAN moves the states to even lower energy.

Although the exchange-splitting is increased with SCAN, it seems like the magnetic pressure associated with a larger magnetic moment, which expands the lattice, is underestimated.
The equilibrium volume of bcc-Fe is reproduced by SCAN at the prize of filled spin-up bands.
For fcc-Ni and hcp-Co, where the spin-up bands are already filled in the LSDA picture, the magnetic moment cannot be much further increased by the increased exchange splitting of SCAN.
This means that the lattice does not expand as much in these systems as for bcc-Fe, resulting in an underestimated equilibrium volume.

The recent calculations of magnetic moments in Fe, Co and Ni by Isaacs and Wolverton \cite{isaacs18} are in very good agreement with our results for the spin moments. 
If the saturation magnetization is compared to the spin magnetic moment, it may appear as if SCAN only slightly overestimates the magnetic moment.
However, if the saturation magnetization is compared to the total magnetic moment it becomes clear that the SCAN functional consistently overestimates the spin moment for all three systems.
Therefore, although SCAN seems to improve the structural equilibrium properties of bcc-Fe, the overall impression is that SCAN does not improve on PBE in itinerant ferromagnets.

We conclude that from the viewpoint of itinerant ferromagnets, further development of accurate exchange-correlation functionals is needed.

\begin{acknowledgments}
\noindent M.\ E.\ and I.\ A.\ are grateful to the Swedish e-Science Research Centre (SeRC) for financial support.
B.\ A.\ gratefully acknowledges financial support by the Swedish Research Council (VR) through the international career grant No. 2014-6336, Marie Sklodowska Curie Actions, Cofund, Project INCA 600398, and the Swedish Foundation for Strategic Research (SSF) through the Future Research Leaders 6 program.

Moreover, we are grateful to the support provided by the Swedish Government Strategic Research Area in Materials Science on Functional Materials at Linköping University (Faculty Grant SFO-Mat-LiU No 2009-00971) and the competence center FunMat-II, financially supported by Vinnova (grant no 2016-05156). The analysis of computational results  was supported by the Russian Science Foundation (project  N\textsuperscript{\underline{o}} 18-12-00492).

All calculations were carried out using the facilities of the Swedish National Infrastructure of Computing (SNIC) at the National Supercomputer Centre (NSC).
\end{acknowledgments}

\end{document}